\documentclass[fleqn]{article}
\usepackage{amsmath}
\usepackage{amsthm}
\usepackage{amssymb}
\usepackage{enumerate}
\usepackage{graphicx}
\theoremstyle{plain}
\newtheorem{lem}{Lemma}
\newtheorem{thm}{Theorem}
\theoremstyle{definition}
\newtheorem*{defn}{Definition}
\theoremstyle{remark}
\newtheorem*{rem}{Remark}

\DeclareMathOperator{\spn}{span}
\DeclareMathOperator{\Aut}{Aut}

\newcommand{\Quad}{Q}

\begin{document}

\Large
\begin{center}
{\bf The Veldkamp space of multiple qubits}
\end{center}
\large
\vspace*{-.1cm}
\begin{center}
P\'eter Vrana and P\'eter L\'evay  
\end{center}
\vspace*{-.4cm} \normalsize
\begin{center}
Department of Theoretical Physics, Institute of Physics, Budapest University of\\ Technology and Economics, H-1521 Budapest, Hungary

\vspace*{.2cm}
(\today)
\end{center}

\vspace*{-.3cm} \noindent \hrulefill

\vspace*{.1cm}
\vspace*{.1cm} \noindent {\bf Abstract}

\noindent
We introduce a point-line incidence geometry in which the commutation relations
of the real Pauli group of multiple qubits are fully encoded. Its points are
pairs of Pauli operators differing in sign and each line contains three pairwise
commuting operators any of which is the product of the other two (up to sign).

We study the properties of its Veldkamp space enabling us to identify subsets of
operators which are distinguished from the geometric point of view. These are
geometric hyperplanes and pairwise intersections thereof.

Among the geometric hyperplanes one can find the set of self-dual operators with
respect to the Wootters spin-flip operation well-known from studies concerning multiqubit entanglement measures. In the two- and three-qubit cases a class of hyperplanes gives rise to
Mermin squares and other generalized quadrangles. 
In the three-qubit case the hyperplane with points corresponding to the $27$ Wootters self-dual operators is just the underlying geometry of the $E_{6(6)}$ symmetric entropy formula describing black holes and strings in five dimensions.

\vspace*{.1cm} \noindent \hrulefill

\section{Introduction}
The importance of generalized Pauli groups in the study of quantum systems with
finite dimensional Hilbert spaces is well known. 
The main application of this group within the field of quantum information is related to quantum error correctiong codes\cite{Nielsen}. 
The construction of such codes is naturally facilitated within the so called stabilizer formalism\cite{Nielsen,Gottesman,Sloane}. 
Here it is recognized that the basic properties of error correcting codes are related to the fact that two operators in the Pauli group are either commuting or anticommuting. This property is encoded into the structure of an Abelian group (the central quotient of the Pauli group), with a natural symplectic structure.
As a next step later it has been realized that  
this commutation algebra for a multiqubit system is
encoded in the totally isotropic subspaces of this underlying  symplectic vector space,
that is, a symplectic polar space of order two \cite{Saniga1}.

Finite geometric concepts in connection with multiqubit Pauli groups also arose in
different contexts, e.g. in connection with discrete phase spaces \cite{Gibbons}, topological quantum computation\cite{Werner} and notably in the
the so-called black hole analogy \cite{LSV,LSVP}. In the latter context, it was shown that there is a
mathematical connection between the Bekenstein-Hawking entropy formula of black
holes and black strings and certain finite
geometric objects related to the three-qubit real Pauli group.
(For a review of the black hole analogy see the paper of Borsten\cite{Borsten} et.al. and references therein.)
More precisely, in a previous paper\cite{LSVP} an explicit connection has been established between the structure of one type of the geometric hyperplanes of the split Cayley hexagon of order two based on the Pauli group for 
three qubits, and the entropy formula for five dimensional black hole and string solutions well-known to string theorists. 
Apart from their use in string theory these studies also emphasized
an important connection between the structure of incidence geometries and their finite automorphism groups realized in terms of quantum gates of quantum information theory\cite{LSV,LSVP}. 
In this spirit groups like the Weyl groups $W(E_6)$ and $W(E_7)$ and the simple group $PSL(2,7)$ as finite subgroups of the infinite discrete U-duality group known from string theory has been linked to the Clifford group of quantum computation\cite{LSV,LSVP,Sole}.

The aim of this paper is twofold. Firstly, we would like to draw the attention
to certain subsets of the $n$-qubit generalized Pauli group which are
distinguished from the finite geometric point of view. These are points and
lines of the Veldkamp space of an incidence geometry naturally associated with
the real Pauli group of $n$ qubits. Some of them have a clear quantum information theoretic  meaning in terms
of the Pauli operators, but for the others this meaning is yet to be found.
Secondly, since these subsets as geometric hyperplanes of our incidence geometry are arising quite naturally also in the black hole analogy we would like to provide a rigorous mathematical frame for these interesting 
constructions.
Such considerations might possibly pave the way for a deeper understanding of this fascinating topic.

The organization of the paper is as follows.
In section 2. we fix our conventions concerning the incidence geometry of the Pauli group. Here we introduce the important notion of a geometric hyperplane.
In section 3. besides providing the basic properties of our incidence geometry, we prove that in this geometry no geometric hyperplane is contained in the other. We also show that a pair of geometric hyperplanes gives naturally rise to a third one.
These considerations lead us in section 4. to initiate a detailed study of the structure of the Veldkamp space, another incidence geometry associated to our initial one,   
with its points being the geometric hyperplanes.
Here we manage to provide an algebraic characterization for the Veldkamp points and establish different relationships between them. 
In section 5. we study the orbits of the action of the symplectic group on the hyperplanes, with the result that there are five types of Veldkamp lines.
In section 6. by investigating the intersection properties of these lines
we manage to obtain a full classification for them.
To put these abstract considerations into physical context we will examine 
some important special cases in section 7.
The conclusions are left for section 8.

\section{The incidence geometry of the Pauli group}
First we briefly summarize the relevant definitions from finite geometry. These
can be found in e.g. \cite{Buek1}. The basic object we will be working with
is the incidence structure whose definition is given here:
\begin{defn}
The triple $(P,L,I)$ is called an incidence structure (or point-line incidence
geometry) if $P$ and $L$ are disjoint sets and $I\subseteq P\times L$ is a
relation. The elements of $P$ and $L$ are called points and lines respectively.
We say that $p\in P$ is incident with $l\in L$ if $(p,l)\in I$.
\end{defn}
The concept of an incidence structure is quite a general one. We may restrict
ourselves to those that are called simple, having the property that no two lines
are incident with exactly the same points. In simple incidence structures the
lines may be identified with the sets of points they are incident with, so we
can think of these as a set $P$ together with a subset $L\subseteq 2^{P}$ of the
power set of $P$. Then $(P,L,\in)$ is an incidence structure in the usual sense.
In what follows we will do this identification, i.e. the points incident with
a line will be called the elements of that line.

In a point-line geometry there are distinguished sets of points called
geometric hyperplanes \cite{Shult}:
\begin{defn}
Let $(P,L,I)$ be an incidence structure. A subset $H\subseteq P$ of $P$ is
called a geometric hyperplane if the following two conditions hold:
\begin{enumerate}[(H1)]
\item $(\forall l\in L):(|H\cap l|=1\textrm{ or }l\subseteq H)$
\item $H\neq P$
\end{enumerate}
\end{defn}

Clearly, the subsets $H$ satisfying only (H1) are exactly the geometric
hyperplanes and the set $P$ of all points.

We will associate a point-line incidence geometry to the generalized real Pauli
group of $n$ qubits for all $n$. This group can be constructed in the following
way. Let us define the $2\times 2$ matrices
\begin{equation}
X=\left[\begin{array}{cc}
0 & 1  \\
1 & 0
\end{array}\right]
\qquad
Z=\left[\begin{array}{cc}
1 & 0  \\
0 & -1
\end{array}\right]
\end{equation}
Observe that these matrices satisfy $X^2=Z^2=I$ where $I$ is the $2\times 2$
identity matrix. The product of the two will be denoted by $Y=ZX=-XZ$. The $n$
qubit real Pauli group is the subgroup of $GL(2^{n},\mathbb{R})$ consisting of
the $n$-fold tensor (Kronecker) products of these four matrices and their
negatives. The shorthand notation $AB\ldots C$ will be used for the tensor product
$A\otimes B\otimes\ldots\otimes C$ of one qubit Pauli group elements $A,B,\ldots,C$,
i.e. we will omit the tensor product sign $\otimes$.  The center of this group
is the same as its commutator subgroup and contains only the identity element
and its negative ($II\ldots I$ and $-II\ldots I$).

It was shown \cite{Sloane,Havlicek} that the central quotient of the Pauli group has the structure of a
symplectic vector space over the field with two elements. The dimension of this
vector space is $2n$, and as the center of the Pauli group contains only the
identity matrix and its negative, the vector addition corresponds to matrix
multiplication up to sign. The symplectic form is induced by the commutator and
has value $0$ if (arbitrary preimages of) the two arguments commute and $1$ if
they anticommute.

Elements of this vector space will be denoted by their
representatives in the Pauli group, for the zero vector we will simply use $0$.
Using the ordered basis consisting of elements containing exactly one $Z$ or $X$
and no $Y$-s we will identify this space with $\mathbb{Z}_{2}^{2n}$ in the following way:
\begin{eqnarray}
I\ldots IIX & \leftrightarrow & (0,0,0,0,\ldots,0,1)  \nonumber  \\
I\ldots IIZ & \leftrightarrow & (0,0,0,0,\ldots,1,0)  \nonumber  \\
 & \vdots & \nonumber  \\
XII\ldots I & \leftrightarrow & (0,1,0,0,\ldots,0,0)  \nonumber  \\
ZII\ldots I & \leftrightarrow & (1,0,0,0,\ldots,0,0)
\end{eqnarray}
In this basis the matrix of the symplectic form is of the form
\begin{equation}
\left[\begin{array}{ccccccc}
0 & 1 & 0 & 0 & \cdots & 0 & 0  \\
1 & 0 & 0 & 0 & \cdots & 0 & 0  \\
0 & 0 & 0 & 1 & \cdots & 0 & 0  \\
0 & 0 & 1 & 0 & \cdots & 0 & 0  \\
\vdots & \vdots & \vdots & \vdots & \ddots & \vdots & \vdots  \\
0 & 0 & 0 & 0 & \cdots & 0 & 1  \\
0 & 0 & 0 & 0 & \cdots & 1 & 0
\end{array}\right]
\end{equation}
Note that we are working in characteristic $2$, therefore every alternating
matrix is also symmetric.
The symplectic form will be denoted by $\langle\cdot,\cdot\rangle$ and the
symplectic vector space $(\mathbb{Z}_{2}^{2n},\langle\cdot,\cdot\rangle)$ by
$V_{n}$.

Recall that the projective space $PG(2n-1,2)$ consists of the nonzero subspaces
of the $2n$ dimensional vector space over the $2$-element field $\mathbb{Z}_2$.
The points of the projective space are one dimensional subspaces of the vector
space, and more generally, $k$ dimensional subspaces of the vector space are
$k-1$ dimensional subspaces of the corresponding projective space.

A subspace of a symplectic vectorspace (and also the subspace in the corresponding
projective space) is called isotropic if there is a vector in it which is orthogonal
to the whole subspace, and totally isotropic if the subspace is orthogonal to
itself. In the case of one and two dimensional (linear) subspaces, the two
notions coincide.

Our incidence geometry consists of the one and two dimensional isotropic subspaces,
i.e. the points and isotropic lines of the projective space $PG(2n-1,2)$. The
collinearity graph of this geometry was studied in \cite{Planat}. Since
the multiplicative group of the invertible elements in the two-element field is
trivial, the points of this projective space can be identified with nonzero vectors.
For later reference we present here the precise definition using our conventions:
\begin{defn}
Let $n\in\mathbb{N}+1$ be a positive integer, and $V_{n}$
be the symplectic $\mathbb{Z}_{2}$-linear space as above.
The incidence structure $\mathcal{G}_{n}$ of the $n$-qubit generalized Pauli
group is $(P,L,\in)$ where $P=V_{n}\setminus\{0\}$,
\begin{equation}
L=\{\{a,b,a+b\}|a,b\in P,a\neq b,\langle a,b\rangle=0\}
\end{equation}
and $\in$ is the set theoretic membership relation.
\end{defn}

In the language of Pauli-operators we can say that points of this incidence
geometry are the pairs of generalized Pauli-operators differing only in a factor
of $-1$ except for the identity element and its negative. On every line there
are three points which are represented by three pairwise commuting operators
any two of which has the third as their product (up to sign).

Our aim will be to find the geometric hyperplanes of the above defined incidence
geometry, and interpret them as special subsets of the real generalized Pauli
group.

\section{Basic properties}

Firstly, we calculate the cardinalities of the point and line sets and the
number of lines incident with one point. Since points in $\mathcal{G}_{n}=(P,L,\in)$
are identified with nonzero vectors of $V_{n}$, it follows that
\begin{equation}\label{eq:pointnum}
|P|=2^{2n}-1=4^{n}-1
\end{equation}
The points collinear with a given point $x$ are
\begin{equation}
C_{x}=\{p\in P|\langle x,p\rangle=0\}=\{p\in V_{n}|\langle x,p\rangle=0\}\setminus\{0\}
\end{equation}
In other words $C_{x}$ is precisely the symplectic complement of the subspace
spanned by $x$ minus the zero vector. Hence, $|C_{x}|=2^{2n-1}-1$. Apart from
$x\in C_{x}$, every element determines a line in $\mathcal{G}_{n}$ passing
through $x$, and every such line is represented by two elements of $C_{x}$.
It follows that the number of lines incident with a given point is $2^{2n-2}-1=4^{n-1}-1$.
The total number of lines is the product of $|P|$ and the latter number divided
by the number of points on a line:
\begin{equation}
|L|=\frac{(4^{n}-1)(4^{n-1}-1)}{3}
\end{equation}

As the next step, we derive some general properties of geometric hyperplanes
of $\mathcal{G}_{n}$. In what follows we exclude $\mathcal{G}_{1}$ from our
consideration in some of the propositions as it is a degenerate case containing
no lines at all. We have the following lower bound on the cardinality of a
geometric hyperplane:
\begin{lem}\label{lem:bound}
Let $n\in\mathbb{N}+2$, $\mathcal{G}_{n}=(P,L,\in)$ and $H\subseteq P$ satisfying (H1). Then the
inequality
\begin{equation}
\frac{|P|}{3}\le|H|
\end{equation}
holds.
\end{lem}
\begin{proof}
A subset $H\subseteq P$ satisfying (H1) must contain at least one point of every
line. Since one point is incident with $4^{n-1}-1$ lines, $|H|$ points can
contain points from at most $|H|(4^{n-1}-1)$ lines. Comparing this with the total
number of lines one obtains
\begin{equation}
|H|(4^{n-1}-1)\ge|L|=\frac{(4^{n}-1)(4^{n-1}-1)}{3}=\frac{|P|(4^{n-1}-1)}{3}
\end{equation}
which implies the statement since $4^{n-1}-1>0$ for $n\ge 2$.
\end{proof}
\begin{rem}
For $n=1$ the statement does not hold: the empty set is a geometric hyperplane
in $\mathcal{G}_{1}$.
\end{rem}

Denoting the number of lines intersecting $H$ in $1$ point by $N_{1}$ and the
number of lines fully contained in $H$ by $N_{2}$ one can write
\begin{equation}
|H|\cdot(4^{n-1}-1)=N_{1}+3N_{2}
\end{equation}
and obviously $N_{1}+N_{2}=|L|$. Solving this system of equations one obtains the
formula
\begin{equation}
N_{2}=\frac{1}{2}(4^{n-1}-1)\left(|H|-\frac{|P|}{3}\right)
\end{equation}
for the number of lines fully containted in a hyperplane. Since this must be
nonnegative, this also yields an alternative proof for the lemma above.

We now give a lower bound on the difference of cardinalities of two subsets
of $P$ satisfying (H1) such that one is contained in the other.
\begin{lem}\label{lem:diff}
Let $n\in\mathbb{N}+2$, $\mathcal{G}_{n}=(P,L,\in)$ and suppose that
$A\subset B\subseteq P$ are two subsets of $P$ satisfying (H1). Then
\begin{equation}
\frac{3}{8}4^{n}\le|B\setminus A|
\end{equation}
\end{lem}
\begin{proof}
For all $p\in P$ let $N_{p}=C_{p}\setminus\{p\}$ denote the set of points
collinear with but not equal to $p$, and for a collinear pair of points $p$ and
$q$ let $N_{pq}=C_{p}\cap C_{q}\setminus\{p,q,p+q\}$ be the set of points
collinear with all of the points of the line containing $p$ and $q$ minus the
points of the line itself. Straightforward calculation shows that $|N_{p}|=2^{2n-1}-2$,
$|N_{x}\cap N_{y}|=2^{2n-2}-3$ and $|N_{pq}|=2^{2n-2}-4$.

Let $x$ be an element of the difference set $B\setminus A$, and pick a line
passing through $x$. Since $A$ and $B$ both satisfy (H1) and $x$ is contained
in $B$ but not in $A$, exactly one of the two other points on this line is in $A$
and the third point $y$ is contained in $B$.

Similar reasoning holds for every line passing through $x$ or $y$, i.e. of
each of these lines exactly one point is contained in $A$ and the whole lines
are contained in $B$. For a point $z\in A$ on one of these lines but not on the line
$\{x,y,x+y\}$ there are three different possibilities. Either $z$ is collinear
with precisely one of $x$ or $y$ or it is collinear with both. These cases are
illustrated on Figure~\ref{fig:cases}. (filled and empty circles correspond to
points of $A$ and $B\setminus A$ respectively).
\begin{figure}[htb]
\centering
\includegraphics{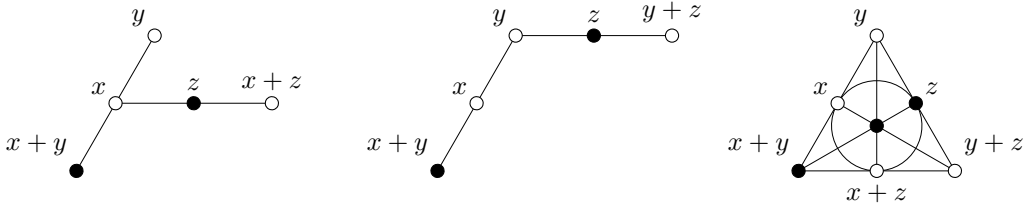}
\caption{The three possible configurations of a point $z$ incident with at least one of two given points $x$ and $y$.\label{fig:cases}}
\end{figure}

In the first two cases $B$ must also contain $x+z$ or $y+z$ respectively. There
are
\begin{equation}
\frac{(|N_{x}\setminus N_{y}|-1)}{2}+\frac{(|N_{y}\setminus N_{x}|-1)}{2}
\end{equation}
such points and they correspond to the same number of points in $B\setminus A$.

In the latter case, when $z$ is collinear with both $x$ and $y$, it is also
collinear with $x+y$, and this implies that the points $\{x,y,z,x+y,x+z,y+z,x+y+z\}$
together with the lines which are subsets of this point-set form a Fano plane.
Then the line $\{x+y,z,x+y+z\}$ is in $A$, and the points $x+z$ and $y+z$ are
outside $A$. Since one such Fano plane contains four points outside the line
$\{x,y,x+y\}$, it follows that the number of them is
\begin{equation}
\frac{|N_{xy}|}{4}
\end{equation}

Each Fano plane containing $\{x,y,x+y\}$ gives rise to two points in $B\setminus A$,
which means that
\begin{eqnarray}
|B\setminus A|
  & \ge & 2+\frac{(|N_{x}\setminus N_{y}|-1)}{2}+\frac{(|N_{y}\setminus N_{x}|-1)}{2}+2\cdot\frac{|N_{xy}|}{4}  \nonumber  \\
  & = & 2+(2^{2n-1}-2)-(2^{2n-2}-3)-1+\frac{2^{2n-2}-4}{2}  \nonumber  \\
  & = & \frac{3}{8}4^{n}
\end{eqnarray}
\end{proof}
\begin{rem}
Again, $n\ge 2$ is needed. In $\mathcal{G}_{1}$, all proper subsets of $P$ are
geometric hyperplanes allowing the difference to consist of a single element,
but $\frac{3}{8}4^{1}=\frac{3}{2}>1$.
\end{rem}

Now we are ready to prove an important fact about the geometric hyperplanes of
$\mathcal{G}_{n}$ ($n\ge 2$).
\begin{thm}\label{thm:Vpoints}
Let $n\in\mathbb{N}+2$, $\mathcal{G}_{n}=(P,L,\in)$ and suppose that
$A,B\subset P$ are two geometric hyperplanes. Then $A\subseteq B$ implies $A=B$,
i.e. no geometric hyperplane is contained in an other one.
\end{thm}
\begin{proof}
Suppose that $A\subset B$ are geometric hyperplanes, one of which is a proper
subset of the other. Then by Lemma~\ref{lem:bound}. we have that
\begin{equation}
\frac{4^{n}-1}{3}\le|A|
\end{equation}
But since $A\subset B\subseteq P$ and $B\subset P\subseteq P$ are two pairs of sets
satisfying the conditions of Lemma~\ref{lem:diff}. we also have that
\begin{eqnarray}
|P|
  & = & |A|+|B\setminus A|+|P\setminus B|  \nonumber  \\
  & \ge & \frac{4^{n}-1}{3}+2\cdot\frac{3}{8}4^{n}  \nonumber  \\
  & = & \frac{13}{12}4^n-\frac{1}{3}
\end{eqnarray}
which contradicts equation (\ref{eq:pointnum}).
\end{proof}

In our incidence geometry every line contains three points. This implies that a
pair of geometric hyperplanes $(A,B)$ gives rise to a third one, the complement of their
symmetric difference which will be denoted by $A\boxplus B$:
\begin{lem}
Suppose that $A\neq B$ are geometric hyperplanes in $\mathcal{G}_{n}=(P,L,\in)$
where $n\ge 1$. Then the set
\begin{equation}
A\boxplus B:=\overline{A\triangle B}=(A\cap B)\cup(\overline{A}\cap\overline{B})=\overline{\overline{A}\triangle \overline{B}}
\end{equation}
(where $\triangle$ denotes the symmetric difference and $\overline{\cdot}=P\setminus\cdot$ is the complement) is also a geometric hyperplane.
\end{lem}
\begin{proof}
Since $A\neq B$, the complement of the symmetric difference is not $P$.

We have to show that given a line $l\in L$, the set $l\cap(A\boxplus B)$ has an
odd number of elements if $l\cap A$ and $l\cap B$ do so.
\begin{eqnarray}
|l\cap(A\boxplus B)|
  & = & |l\cap A\cap B|+|l\cap(P\setminus(A\cup B))|  \nonumber  \\
  & = & |l\cap A\cap B|+|l\setminus(A\cup B)|
\end{eqnarray}
If any of $A$ and $B$ contains $l$ then the latter term is $0$, and the first
term equals $|l\cap A|$ or $|l\cap B|$ both of which are odd.

If both of $A$ and $B$ meet $l$ in a single point then either these points
coincide or they are different points of $l$. In the first case the first term
is $1$ and the second is $2$, and in the second case the first term equals $0$
and the second is $|l|-2=1$.
\end{proof}

A simple observation about this operation is that
\begin{eqnarray}
A\cap(A\boxplus B)
  & = & A\cap((A\cap B)\cup(\overline{A}\cap\overline{B})  \nonumber  \\
  & = & A\cap B
\end{eqnarray}
and similarly for $B\cap(A\boxplus B)$. Moreover, any two of the triple
$\{A,B,A\boxplus B\}$ determines the third one, since
\begin{eqnarray}
A\boxplus(A\boxplus B)
  & = & \overline{A\triangle\overline{A\triangle B}}  \nonumber  \\
  & = & \overline{\overline{A}\triangle(\overline{A}\triangle\overline{B})}  \nonumber  \\
  & = & B
\end{eqnarray}
In fact, $\boxplus$ makes the set of subsets of $P$ satisfying (H1) a
$\mathbb{Z}_{2}$-linear space with the set $P$ as the zero vector.

\section{The Veldkamp space}

To certain incidence geometries one can associate another inciedence geometry
called its Veldkamp space whose points are geometric hyperplanes \cite{Shult}:
\begin{defn}
Let $\Gamma=(P,L,I)$ be a point-line geometry. We say that $\Gamma$ has
Veldkamp points and Veldkamp lines if it satisfies the conditions
\begin{enumerate}[(V1)]
\item For any hyperplane $A$ it is not properly contained in any other hyperplane $B$.
\item For any three distinct hyperplanes $A$, $B$ and $C$, $A\cap B\subseteq C$ implies $A\cap B=A\cap C$.
\end{enumerate}

If $\Gamma$ has Veldkamp points and Veldkamp lines, then we can form the
Veldkamp space $V(\Gamma)=(P_{V},L_{V},\supseteq)$ of $\Gamma$, where $P_{V}$ is
the set of geometric hyperplanes of $\Gamma$, and $L_{V}$ is the set of
intersections of pairs of distinct hyperplanes.
\end{defn}

By Theorem \ref{thm:Vpoints}., the incidence geometry $\mathcal{G}_{n}$ has
Veldkamp points for $n\ge 2$. Now we give the explicit form of geometric
hyperplanes. To this end let us introduce a quadratic form over $V_{n}$
whose linearization is the symplectic form $\langle\cdot,\cdot\rangle$:
\begin{equation}
\Quad_{0}(x)=\sum_{i=1}^{n}a_{i}b_{i}
\end{equation}
where $x=(a_1,b_1,a_2,b_2,\ldots,a_n,b_n)\in V_{n}$. It is easy to
check that
\begin{equation}
\Quad_{0}(x+y)+\Quad_{0}(x)+\Quad_{0}(y)=\langle x,y\rangle
\end{equation}
It is also true that $\Quad_{0}(x)=0$ iff the Pauli operators representing $x$
are symmetric matrices. To every element $p$ in $V_{n}$ we can associate a
nondegenerate quadratic form
\begin{equation}
\Quad_{p}(x)=\Quad_{0}(x)+\langle p,x\rangle
\end{equation}
whose linearized form is the same as that of $\Quad_{0}$:
\begin{eqnarray}
\Quad_{p}(x)+\Quad_{p}(y)+\Quad_{p}(x+y)
  & = & \Quad_{0}(x)+\langle p,x\rangle+\Quad_{0}(y)+\langle p,y\rangle+  \nonumber  \\
  & & +\Quad_{0}(x+y)+\langle p,x+y\rangle  \nonumber  \\
  & = & \Quad_{0}(x)+\Quad_{0}(y)+\Quad_{0}(x+y)+  \nonumber  \\
  & & +\langle p,x\rangle+\langle p,y\rangle+\langle p,x+y\rangle  \nonumber  \\
  & = & \langle x,y\rangle
\end{eqnarray}
We will use these quadratic forms to characterize points of geometric hyperplanes.

An important concept is the Arf invariant of a quadratic form over a
$\mathbb{Z}_2$-linear space which is the element of $\mathbb{Z}_2$ that occurs
most often among the values of the form. It is not hard to check that the Arf
invariant of $\Quad_{x}$ is $\Quad_{0}(x)$ where $\Quad_{0}$ is a quadratic
form with Arf invariant 0.

\begin{lem}\label{lem:hypplanes}
Let $n\in\mathbb{N}+1$ be a positive integer, $\mathcal{G}_{n}=(P,L,\in)$ and
$p\in V_{n}$ be any vector. Then the sets
\begin{equation}
C_{p}=\{x\in P|\langle p,x\rangle=0\}
\end{equation}
and
\begin{equation}
H_{p}=\{x\in P|\Quad_{p}(x)=0\}
\end{equation}
satisfy (H1).
\end{lem}
\begin{proof}
Since $C_{p}$ is a projective subspace of codimension $1$ (or $0$ in the case of
$x=0$) in $PG(2n-1,2)$ it intersects every line in this projective space (not
only the isotropic ones) in either $1$ or $3$ points.

Now let $l=\{a,b,a+b\}\in L$ be a line in $\mathcal{G}_{n}$. Since it has $3$
points, we only have to show that $\overline{H_{p}}$ intersects $l$ in an even
number of points. This is implied by
\begin{equation}
\Quad_{p}(a)+\Quad_{p}(b)+\Quad_{p}(a+b)=\langle a,b\rangle=0
\end{equation}
\end{proof}

Clearly, $C_{0}=P$ but all other sets appearing in Lemma~\ref{lem:hypplanes}. are
geometric hyperplanes. In fact, the converse is also true, i.e. all geometric
hyperplanes arise in this form:
\begin{thm}
Let $n\in\mathbb{N}+1$, $\mathcal{G}_{n}=(P,L,\in)$, and $H\in P$ a subset
satisfying (H1). Then either $H=C_{p}$ or $H=H_{p}$ for some $p\in V_{n}$.
\end{thm}
\begin{proof}
We prove by induction. For $n=1$ one can check that the $8=2\cdot 2^{2}$ possible
subsets of $P$ are indeed of this form. For $n\ge 2$ we can write $n$ as the
sum of two positive integers $a$ and $b$. Then $V_{n}\simeq V_{a}\oplus V_{b}$
and the points of $\mathcal{G}_{n}$ are
\begin{multline}
P=\{p_{a}\oplus 0|p_{a}\in V_{a}\setminus\{0\}\}\cup\{0\oplus p_{b}|p_{b}\in V_{b}\setminus\{0\}\}  \\
  \cup\{p_{a}\oplus p_{b}|p_{a}\in V_{a}\setminus\{0\},p_{b}\in V_{b}\setminus\{0\}\}
\end{multline}
The first set in the union will be denoted by $P_{a}$, the second one by $P_{b}$,
and the last one can naturally be identified with $P_{a}\times P_{b}$.
The first two sets can be regarded as the point-sets of $\mathcal{G}_{a}$ and
$\mathcal{G}_{b}$ respectively. The latter two incidence structures arise as
these points and the lines contained in the appropriate point-sets.

Now let $H^{(a)}=H\cap P_{a}$ and $H^{(b)}=H\cap P_{b}$. Clearly, they satisfy
(H1) in $\mathcal{G}_{a}$ and $\mathcal{G}_{b}$, so by the induction hypothesis
they are either of the form $C_{p_{a}}$ ($C_{p_{b}}$) or $H_{p_{a}}$ ($H_{p_{b}}$)
for some $p_{a}\in V_{a}$ ($p_{b}\in V_{b}$). In any case, since every point in
$P_{a}$ is connected with every point in $P_{b}$ the points in the intersections
uniquely determine the set $H$:
\begin{equation}
H=H^{(a)}\cup H^{(b)}\cup(H^{(a)}\times H^{(b)})\cup((P_{a}\setminus H^{(a)})\times(P_{b}\setminus H^{(b)}))
\end{equation}

We have the following three cases (after possibly reversing the role of $a$ and $b$):
\begin{enumerate}[a)]
\item $H^{(a)}=C_{p_{a}}$ and $H^{(b)}=C_{p_{b}}$.
Then
\begin{eqnarray}
H
  & = & \{x\oplus 0|\langle p_{a},x\rangle=0\}\cup\{0\oplus y|\langle p_{b},y\rangle=0\}  \nonumber  \\
  & & \cup\{x\oplus y|\langle p_{a},x\rangle=\langle p_{b},y\rangle=0\}\cup\{x\oplus y|\langle p_{a},x\rangle=\langle p_{b},y\rangle=1\}  \nonumber  \\
  & = & \{x\oplus y|\langle p_{a},x\rangle+\langle p_{b},y\rangle=0\}  \nonumber  \\
  & = & \{x\oplus y|\langle p_{a}\oplus p_{b},x\oplus y\rangle=0\}  \nonumber  \\
  & = & C_{p_{a}\oplus p_{b}}
\end{eqnarray}

\item $H^{(a)}=H_{p_{a}}$ and $H^{(b)}=H_{p_{b}}$.
Then
\begin{eqnarray}
H
  & = & \{x\oplus 0|\Quad_{p_{a}}(x)=0\}\cup\{0\oplus y|\Quad_{p_{b}}(y)=0\}  \nonumber  \\
  & & \cup\{x\oplus y|\Quad_{p_{a}}(x)=\Quad_{p_{b}}(y)=0\}\cup\{x\oplus y|\Quad_{p_{a}}(x)=\Quad_{p_{b}}(y)=1\}  \nonumber  \\
  & = & \{x\oplus y|\Quad_{p_{a}}(x)+\Quad_{p_{b}}(y)=0\}  \nonumber  \\
  & = & \{x\oplus y|\Quad_{p_{a}\oplus p_{b}}(x\oplus y)=0\}  \nonumber  \\
  & = & H_{p_{a}\oplus p_{b}}
\end{eqnarray}

\item $H^{(a)}=H_{p_{a}}$ and $H^{(b)}=C_{p_{b}}$.
In this case $H$ would be
\begin{equation}
H=\{x\oplus y|\Quad_{p_{a}}(x)+\langle p_{b},y\rangle=0\}
\end{equation}
Since $a\ge 1$ and $b\ge 1$ we can pick from $V_{a}$ and $V_{b}$ two dimensional
symplectic subspaces $W_{1}$ and $W_{2}$ which are direct summands in the appropriate
subspaces. Then $p_{a}$ ($p_{b}$) can be uniquely written as a sum of a vector
$p_{1}$ ($p_{2}$) in $W_{1}$ ($W_{2}$) and one in its direct complement.
Clearly, $H$ intersects $W_{1}\oplus W_{2}$ in
\begin{equation}
H\cap(W_{1}\oplus W_{2})=\{x_{1}\oplus x_{2}|\Quad_{p_{1}}(x_{1})+\langle p_{2},x_{2}\rangle=0\}
\end{equation}
and this set should be a geometric hyperperplane in the $\mathcal{G}_{2}$ whose
points are $P\cap(W_{1}\oplus W_{2})$. But denoting the three points in
$P\cap W_{i}$ with $a_{i},b_{i},c_{i}$ ($i\in\{1,2\}$) one can write that
\begin{equation}
\langle a_{1}\oplus a_{2},b_{1}\oplus b_{2}\rangle=\langle a_{1},b_{1}\rangle+\langle a_{2},b_{2}\rangle=1+1=0
\end{equation}
and
\begin{equation}
(a_{1}\oplus a_{2})+(b_{1}\oplus b_{2})=(c_{1}\oplus c_{2})
\end{equation}
so these points form a line in $\mathcal{G}_{2}$, and
\begin{multline}
\Quad_{p_{1}}(a_{1})+\langle p_{2},a_{2}\rangle+\Quad_{p_{1}}(b_{1})+\langle p_{2},b_{2}\rangle+\Quad_{p_{1}}(c_{1})+\langle p_{2},c_{2}\rangle  \\
  = \langle a_{1},b_{1}\rangle+\langle p_{2},a_{2}+b_{2}+(a_{2}+b_{2})\rangle=1
\end{multline}
shows that an even number of points in this line belong to $H$ which contradicts
(H1).
\end{enumerate}
\end{proof}

Given this algebraic characterization of points of different hyperplanes it is
easy to express the sum $A\boxplus B$ of any two geometric hyperplanes $A$ and
$B$:
\begin{lem}\label{lem:Vlines}
Let $n\in\mathbb{N}+1$, $a,b\in V_{n}$ and $\mathcal{G}_{n}=(P,L,\in)$. Then the
following formulas hold:
\begin{eqnarray}
C_{a}\boxplus C_{b} & = & C_{a+b}  \nonumber  \\
H_{a}\boxplus H_{b} & = & C_{a+b}  \\
C_{a}\boxplus H_{b} & = & H_{a+b}  \nonumber
\end{eqnarray}
\end{lem}
\begin{proof}
Since all of the arising sets are defined as the zero locus of some $\mathbb{Z}_{2}$-valued
function, we only have to observe that
\begin{equation}
\{x\in P|f_{1}(x)=0\}\boxplus\{x\in P|f_{2}(x)=0\}=\{x\in P|f_{1}(x)+f_{2}(x)=0\}
\end{equation}
holds.

Since $\langle a,x\rangle+\langle b,x\rangle=\Quad_{a}(x)+\Quad_{b}(x)=\langle a+b,x\rangle$
and $\langle a,x\rangle+\Quad_{b}(x)=\Quad_{0}(x)+\langle a,x\rangle+\langle b,x\rangle=\Quad_{a+b}(x)$,
the statement follows.
\end{proof}

Now we are ready to prove that condition (V2) holds:
\begin{thm}
Let $n\in\mathbb{N}+3$, and suppose that $A,B,C$ are distinct geometric
hyperplanes of $\mathcal{G}_{n}=(P,L,\in)$ such that $I=A\cap B\subseteq C$.
Then $A\cap B=A\cap C$.
\end{thm}
\begin{proof}
For $C=A\boxplus B$ we have seen that $A\cap B=A\cap C=B\cap C$. We will show
that there is no other possibility, i.e. $A\cap B\subseteq C$ implies $C\in\{A,B,A\boxplus B\}$.

By Lemma~\ref{lem:Vlines}. and the properties of $\boxplus$, we may assume that
$A=C_{a}$ for some $a\in P$. We have two possibilities:
\begin{enumerate}[a)]
\item $B=C_{b}$ for some $b\in P$.
Then $I=\{x\in P|\langle a,x\rangle=\langle b,x\rangle=0\}$, so
$I\cup\{0\}=(\spn\{a,b\})^{\perp}$ is the symplectic complement of the subspace
spanned by $a$ and $b$. This means that $I\subseteq C_{v}=v^{\perp}\setminus\{0\}$
implies $v\in\spn\{a,b\}$, or in other words, $C_{v}$ is one of $A$, $B$ and $A\boxplus B$.

In order to show that we cannot find a $v\in V_{n}$ such that $I\subseteq H_{v}$,
pick two points $x,y\in I$ such that $\langle x,y\rangle=1$. It is possible
since $\dim V_{n}\ge 6$ when $n\ge 3$, $\dim\spn I=\dim V_{n}-2>\frac{\dim V_{n}}{2}$
therefore $\spn I$ cannot be isotropic. Then $x+y$ is also contained in $I$ because
$I\cup\{0\}$ is a linear subspace in $V_{n}$ and $x\neq y$.
\begin{equation}
\Quad_{v}(x)+\Quad_{v}(y)+\Quad_{v}(x+y)=\langle x,y\rangle=1
\end{equation}
shows that $\{x,y,x+y\}$ cannot be contained in $H_{v}$.

\item $B=H_{b}$ for some $b\in V_{n}$.
Then $\Quad_{b}$ restricted to $a^{\perp}$ is a quadratic form of maximal rank,
and $I\cup\{0\}$ is its zero locus. Therefore $\spn I=a^{\perp}$ and it follows that
if $I\subseteq C_{v}=v^{\perp}\setminus\{0\}$ then $v\in I^{\perp}=\spn\{a\}$.

If for some $v\in V_{n}$ $I\subseteq H_{v}$, then
\begin{equation}
I=I\cap H_{b}\subseteq H_{v}\cap H_{b}\subseteq H_{v}\boxplus H_{b}=C_{v+b}
\end{equation}
which means that the only possibilities are $v=b$ and $v=a+b$.
\end{enumerate}
\end{proof}

\begin{rem}
The statement is not true for $n=2$. There the perp-sets of two commuting
operators intersect in a single line which is obviously contained in a grid
whose intersection with any of the two given perp-sets is a pentad \cite{Buek2,Saniga2}.
\end{rem}

\section{Automorphisms}
Now that we have characterized all the geometric hyperplanes of $\mathcal{G}_n$,
it is convenient to calculate how do automorphisms of $\mathcal{G}_n$ act on
them. It is clear that every automorphism of $V_n$ (i.e. a symplectic
transformation) induces one of $\mathcal{G}_n$ and this group homomorphism is
injective. Conversely, an automorphism of $\mathcal{G}_n$ respects the linear
structure by Lemma~\ref{lem:Vlines}., and preserves the symplectic structure too,
since it maps lines to lines. It follows then, that $\Aut(\mathcal{G}_n)=Sp(2n,2)$.

We have three types of geometric hyperplanes. One of them is $C_{p}$ where
$p\in V_{n}\setminus\{0\}$, and the two other types are of the form $H_{p}$ where
$p\in V_{n}$. The type of this depends on the Arf invariant of $\Quad_p$ which
in turn equals $\Quad_0(p)$ which we will also call the Arf invariant of the
hyperplane. The number of hyperplanes of each type is summarized in Table~\ref{tab:Vpoints}.

\begin{table}
\centering
\begin{tabular}{c|c|c}
general form  &  number of points  &  copies  \\
\hline
$C_{p}$ where $p\in V_{n}\setminus\{0\}$  &  $\frac{1}{2}4^{n}-1$  &  $4^n-1$  \\
$H_{p}$ where $\Quad_{0}(p)=0$  &  $\frac{1}{2}(4^{n}+2^{n})-1$  &  $\frac{1}{2}(4^{n}+2^{n})$  \\
$H_{p}$ where $\Quad_{0}(p)=1$  &  $\frac{1}{2}(4^{n}-2^{n})-1$  &  $\frac{1}{2}(4^{n}-2^{n})$
\end{tabular}
\caption{Geometric hyperplanes in the incidence geometry associated to the $n$-qubit Pauli group\label{tab:Vpoints}}
\end{table}

Let $\mathcal{G}_n=(P,L,\in)$ where $P=V_{n}\setminus\{0\}$. The action of $Sp(2n,2)$
on $V_{n}$ induces an action on the Veldkamp space of $\mathcal{G}_{n}$. Since
$Sp(2n,2)$ is generated by symplectic transvections, we only have to calculate
the action of these on the set of geometric hyperplanes. Let $t_{p}$ denote
the symplectic transvection determined by $p$:
\begin{equation}
t_{p}:V_{n}\to V_{n};\quad x\mapsto x+\langle p,x\rangle p
\end{equation}
A well-known fact is that the inverse of $t_{p}$ is itself, that is, $t_{p}$ is
an involution.

An other important property is that if $\langle p,q\rangle=0$ then $t_{p}$ and
$t_{q}$ commute:
\begin{eqnarray}
t_{p}t_{q}x
  & = &  t_{p}(x+\langle q,x\rangle q)  \nonumber  \\
  & = &  x+\langle q,x\rangle q+\langle p,x+\langle q,x\rangle q\rangle p  \nonumber  \\
  & = &  x+\langle q,x\rangle q+\langle p,x\rangle p+\langle q,x\rangle\langle p,q\rangle p
\end{eqnarray}
$\langle p,q\rangle=0$ means that the last term is zero and the rest is symmetric
in $p$ and $q$.

Since $t_{p}$ is linear, in particular, it fixes the zero vector, it acts as a
permutation of $P$ too. Being symplectic it permutes the elements of $L$ too,
and maps hyperplanes to hyperplanes. The action on these is given by
\begin{lem}
Let $\mathcal{G}_{n}=(P,L,\in)$ and $p\in V_{n}$. Then
\begin{eqnarray}\label{eq:action}
t_{p}C_{a} & = & C_{t_{p}a}  \nonumber  \\
t_{p}H_{a} & = & H_{a+(1+\Quad_{a}(p))p}
\end{eqnarray}
\end{lem}
\begin{proof}
\begin{eqnarray}
t_{p}C_{a}
  & = & \{t_{p}x|x\in V_{n}\setminus\{0\},\langle a,x\rangle=0\}  \nonumber  \\
  & = & \{x\in P|\langle a,t_{p}x\rangle=0\}  \nonumber  \\
  & = & \{x\in P|\langle t_{p}a,x\rangle=0\}  \nonumber  \\
  & = & C_{t_{p}a}
\end{eqnarray}
\begin{eqnarray}
t_{p}H_{a}
  & = & \{t_{p}x|x\in V_{n}\setminus\{0\},\Quad_{a}(x)=0\}  \nonumber  \\
  & = & \{x\in P|\Quad_{a}(t_{p}x)=0\}  \nonumber  \\
  & = & \{x\in P|\Quad_{a}(x+\langle p,x\rangle p)=0\}  \nonumber  \\
  & = & \{x\in P|\Quad_{0}(x)+\langle p,x\rangle\Quad_{0}(p)+\langle x,\langle p,x\rangle p\rangle+\langle a,x+\langle p,x\rangle p\rangle=0\}  \nonumber  \\
  & = & \{x\in P|\Quad_{0}(x)+\langle\Quad_{0}(p)p+a+\langle a,p\rangle p+p,x\rangle=0\}  \nonumber  \\
  & = & H_{a+(1+\Quad_{0}(p)+\langle a,p\rangle)p}
\end{eqnarray}
\end{proof}
\begin{rem}
In particular, $t_{p}$ fixes $C_{a}$ iff $\langle p,a\rangle=0$ and fixes $H_{a}$
iff $\Quad_{a}(p)=1$.
\end{rem}

It is well known that $Sp(2n,2)$ acts transitively on the set of pairs of distinct
nonzero vectors in $V_{n}$ with fixed symplectic product. This means that two
hyperplanes of type $C_{p}$ can be in two different positions relative to each
other.

Our aim is to identify the possible relative positions of two geometric
hyperplanes of type $H_{p}$. Clearly, the set
\begin{equation}
\{\{H_{a},H_{b}\}|a,b\in V_{n},a\neq b\}
\end{equation}
splits to at least three invariant subsets under the action of $Sp(2n,2)$, namely
\begin{eqnarray}
\{\{H_{a},H_{b}\}|a,b\in V_{n},a\neq b,\Quad_{0}(a)=\Quad_{0}(b)=0\}  \\
\{\{H_{a},H_{b}\}|a,b\in V_{n},a\neq b,\Quad_{0}(a)=\Quad_{0}(b)=1\}  \\
\{\{H_{a},H_{b}\}|a,b\in V_{n},\Quad_{0}(a)\neq\Quad_{0}(b)\}
\end{eqnarray}

We will show that $Sp(2n,2)$ acts transitively on each of these sets. This
follows from the following lemma:
\begin{lem}\label{lem:transitive}
Let $n\in\mathbb{N}+3$, $\mathcal{G}_n=(P,L,\in)$ and $a,b,f\in V_{n}$ three
distinct vectors such that $\Quad_{0}(a)=\Quad_{0}(b)$. Then there exists
an element in $Sp(2n,2)$ fixing $H_{f}$ and swapping $H_{a}$ with $H_{b}$.
\end{lem}
\begin{proof}
There are two possibilities according to the value of $\Quad_{f}(a+b)$.
\begin{enumerate}[a)]
\item 
If $\Quad_{f}(a+b)=0$, then pick a point $p$ in $C_{a+b}\cap H_{a}\setminus H_{f}$.
This is possible since $C_{a+b}\boxplus H_{a}=H_{b}\neq H_{f}$ and (V2) holds.
Now let $q=a+b+p$. Then since $H_{f}$ is a geometric hyperplane, $a+b\in H_{f}$
and $p\notin H_{f}$, it follows that the line $\{q,p,a+b\}$ intersects $H_{f}$
in $a+b$ and $\Quad_{f}(q)=1$.

Also, we have that
\begin{eqnarray}
\Quad_{b}(a+b+p)
  & = &  \Quad_{0}(a+b+p)+\langle b,a+b+p\rangle  \nonumber  \\
  & = &  \Quad_{0}(a)+\Quad_{0}(b)+\Quad_{0}(p)+\langle a,p\rangle  \nonumber  \\
  & = &  \Quad_{a}(p)=0
\end{eqnarray}

It is clear then by equation~(\ref{eq:action}), that both $t_{p}$ and $t_{q}$ fixes $H_{f}$, $t_{p}H_{a}=H_{a+p}$
and $t_{q}H_{b}=H_{b+q}=H_{a+p}$. It follows that $t_{q}t_{p}H_{a}=H_{b}$. Since
$\langle p,q\rangle=0$, $t_{p}$ and $t_{q}$ are two commuting involutions, which
implies that $t_{q}t_{p}$ itself is an involution and it swaps $H_{a}$ and $H_{b}$

\item
If $\Quad_{f}(a+b)=1$ then by equation~(\ref{eq:action}), $t_{a+b}$ fixes $H_{f}$ and
\begin{equation}
\Quad_{a}(a+b) = \Quad_{0}(a)+\Quad_{0}(b)+\langle a,b\rangle+\langle a,a+b\rangle = 0
\end{equation}
implies that $t_{a+b}H_{a}=H_{b}$.
\end{enumerate}
\end{proof}
\begin{rem}
This means also that $Sp(2n,2)$ acts 2-transitively on its two orbits of
geometric hyperplanes of type $H$.
\end{rem}

\section{Veldkamp lines}
Our considerations in the previous section show that there are two types of
Veldkamp lines incident with three $C$-hyperplanes and three types of lines
which are incident with one $C$-hyperplane and two $H$-hyperplanes. In this
section we study the structure of these lines, i.e. the pairwise intersections
of geometric hyperplanes.

By Lemma~\ref{lem:transitive}. we only have to consider five special cases. The
first type of line in the Veldkamp space of $\mathcal{G}_n$ we study is the one
connecting $C_a$ and $C_b$ where $\langle a,b\rangle=0$. Their intersection
contains points of the symplectic complement of $\spn\{a,b\}$. Since this
subspace is isotropic, the symplectic complement is isomorphic to $V_{n-2}\oplus W_{1}$
where $W_{1}$ is a 2-dimensional vector space with an identically 0 bilinear
form. Therefore the intersection has $4^{n-1}-1$ points. The number of lines of
this type is the same as the number of lines in $\mathcal{G}_n$.

Similarly, when $\langle a,b\rangle=1$, then the intersection of $C_a$ and $C_b$
is also the symplectic complement of $\spn\{a,b\}$ minus the zero vector. But
in this case the symplectic form restricted to the complement is nondegenerate,
meaning that the intersection as an incidence geometry is isomorphic to $\mathcal{G}_{n-1}$
which has $4^{n-1}-1$ points. Each two dimensional symplectic subspace in $V_{n}$
gives rise to one such line, hence the number of them is
\begin{equation}
\frac{(4^{n}-1)\cdot 4^{n-1}}{3}
\end{equation}

The next case is a line connecting two $H$-hyperplanes with Arf invariants equal
to $0$. By Lemma~\ref{lem:transitive}., we can choose any two such hyperplanes. For
simplicity we will work with $H_{0}$ and $H_{a}$ where $a=I\ldots IX=(0,0,\ldots,0,1)$. Now
$x\in H_{0}\cap H_{a}$ implies that $x\in C_{a}$, so $x$ is of the form
\begin{equation}
x=(x_{1},x_{2},\ldots,x_{2n-2},0,x_{2n})
\end{equation}
The value of $\Quad_{0}(x)$ is independent of $x_{2n}$ and equals
\begin{equation}
\Quad_{0}((x_{1},x_{2},\ldots,x_{2n-2}))=x_{1}x_{2}+\ldots+x_{2n-3}x_{2n-2}
\end{equation}
where $\Quad_{0}$ denotes also a quadratic form on $V_{n-1}$, but this should
not be a source of confusion. As we are free to choose the value of $x_{2n}$,
it follows that the intersection contains
\begin{equation}
2\left[\frac{1}{2}(4^{n-1}+2^{n-1})\right]-1=4^{n-1}+2^{n-1}-1
\end{equation}
points where $-1$ is for excluding the zero vector. We have
\begin{eqnarray}
\binom{\frac{4^n+2^n}{2}}{2}
  & = &  \frac{1}{8}(4^n+2^n)(4^n+2^n-2)  \nonumber  \\
  & = &  2^{n-3}(2^n+1)(2^n-1)(2^n+2)  \nonumber  \\
  & = &  2^{n-3}(4^n-1)(2^n+2)
\end{eqnarray}
lines of this type.

For the line type connecting two $H$-hyperplanes with different Arf invariants
we choose $H_{0}$ and $H_{a}$, $a=I\ldots IY=(0,0,\ldots,0,1,1)$ as representative. Now if
$x\in H_{0}\cap H_{a}$ then $x\in C_{a}$, so $x=(x_{1},x_{2},\ldots,x_{2n-2},x_{2n-1},x_{2n-1})$.
This implies that $\Quad_{0}(x)$ equals
\begin{equation}
\Quad_{0}((x_{1},x_{2},\ldots,x_{2n-2}))+x_{2n-1}
\end{equation}
meaning that we can choose the first $2n-2$ coordinates freely, and this uniquely
determines $x_{2n-1}$. Moreover, since the last two coordinates do not contribute
to the value of the symplectic form of two such vectors, it follows that the
intersection is again isomorphic to $\mathcal{G}_{n-1}$ as an incidence geometry.
But this time it is embedded differently into $\mathcal{G}_n$ since the span
of its points is $2n-1$ dimensional unlike the symplectic complement of a symplectic
2-dimensional subspace whose span is $2n-2$ dimensional.

A pair of $H$-hyperplanes with different Arf invariants can be chosen in
\begin{eqnarray}
\frac{1}{2}(4^n+2^n)\cdot\frac{1}{2}(4^n-2^n)
  & = & \frac{1}{4}(2^n)^2(2^n+1)(2^n-1)  \nonumber  \\
  & = & 4^{n-1}(4^n-1)
\end{eqnarray}
ways.

The last case is a line containing two $H$-hyperplanes with Arf invariant 1.
Our choice is $a=I\ldots IY=(0,0,\ldots,0,1,1)$ and $b=I\ldots IXY=(0,\ldots,0,1,1,1)$, and the two
hyperplanes are $H_{a}$ and $H_{b}$. Then for $x\in H_{a}\cap H_{b}$, we have that
$x\in C_{a+b}$ meaning that $x=(x_{1},x_{2},\ldots,x_{2n-4},0,x_{2n-2},x_{2n-1},x_{2n})$
Now $\Quad_{a}$ equals
\begin{equation}
\Quad_{0}((x_{1},x_{2},\ldots,x_{2n-4}))+x_{2n-1}+x_{2n}+x_{2n-1}x_{2n}
\end{equation}
which does not depend on $x_{2n-2}$. If any of the last two
coordinates is $1$, then the sum of the last three terms is $1$, so we have
$\frac{1}{2}(4^{n-2}-2^{n-2})$ possibilities for the values of the first $n-4$
coordinates. On the other hand, if both $x_{2n-1}$ and $x_{2n}$ are $0$, then
we have $\frac{1}{2}(4^{n-2}+2^{n-2})$ choices for the first $n-4$ coordinates.
In any case, we are free to choose the value of $x_{2n-2}$ except for the case
of $x=0$, so the intersection has
\begin{equation}
2\left(3\frac{1}{2}(4^{n-2}-2^{n-2})+\frac{1}{2}(4^{n-2}+2^{n-2})\right)-1=4^{n-1}-2^{n-1}-1
\end{equation}
points. The number of lines of this type is
\begin{eqnarray}
\binom{\frac{4^n-2^n}{2}}{2}
  & = &  \frac{1}{8}(4^n-2^n)(4^n-2^n-2)  \nonumber  \\
  & = &  2^{n-3}(2^n-1)(2^n+1)(2^n-2)  \nonumber  \\
  & = &  2^{n-3}(4^n-1)(2^n-2)
\end{eqnarray}

The above results are summarized in Table~\ref{tab:Vlines}.

\begin{table}
\centering
\begin{tabular}{c c c|c|c}
\multicolumn{3}{c}{hyperplanes of type}  &  intersection  &  number of  \\
$C_{a}$  &  $H_{a}$, $\Quad_0(a)=0$  &  $H_{a}$, $\Quad_0(a)=1$  &  size  &  copies  \\
\hline
$3$  &  $0$  &  $0$  &  $4^{n-1}-1$  &  $\frac{1}{3}(4^{n}-1)(4^{n-1}-1)$  \\
$3$  &  $0$  &  $0$  &  $4^{n-1}-1$  &  $\frac{1}{3}4^{n-1}(4^{n}-1)$  \\
$1$  &  $2$  &  $0$  &  $4^{n-1}+2^{n-1}-1$  &  $2^{n-3}(4^n-1)(2^n+2)$ \\
$1$  &  $1$  &  $1$  &  $4^{n-1}-1$  &  $4^{n-1}(4^n-1)$ \\
$1$  &  $0$  &  $2$  &  $4^{n-1}-2^{n-1}-1$  &  $2^{n-3}(4^n-1)(2^n-2)$ \\
\end{tabular}
\caption{Veldkamp lines in the incidence geometry associated to the $n$-qubit Pauli group\label{tab:Vlines}}
\end{table}

\section{Special cases}
After the discussion of general properties of the incidence geometries of the
$n$-qubit Pauli group, we turn to some special cases. As it was already
mentioned, $\mathcal{G}_{1}$ consists of three points and no lines. This is
not very interesting, as all proper subsets arise as geometric hyperplanes and
intersections of those are all subsets with at most one point.

The $n=2$ case is rather peculiar as $\mathcal{G}_{2}$ is the unique generalized
quadrangle of order two having $15$ points and $15$ lines. It was already
studied in detail in \cite{Saniga2,Saniga3}. Our present results can be regarded as a generalization
of this case. It is interesting to note that in this case an $H$-hyperplane with
Arf invariant $1$ consists of $5$ points reaching the lower bound of Lemma~\ref{lem:bound}.
These do not have any lines, and hence are ovoids, corresponding
to sets of mutually anticommuting Pauli operators.

The $H$-hyperplanes with Arf invariant $0$ contain $9$ points and $6$ lines
forming a subquadrangle $GQ(2,1)$ also known as a grid. On the quantum information
theoretic side these are Mermin squares which are used for a simplified proof
of the Kochen-Specker theorem \cite{Mermin}.

For $n=3$ the incidence geometry $\mathcal{G}_{n}$ has $63$ points and $315$
lines. The $H$-hyperplanes with Arf invariant $1$ here have $27$ points and
$45$ lines. These hyperplanes as incidence geometries are isomorphic to the
generalized quadrangle $GQ(2,4)$. Fixing one of them, all of its geometric hyperplanes
can be obtained by intersecting it with every other hyperplane in $\mathcal{G}_{3}$.
They are $GQ(2,2)$-s and perp-sets containing $11$ points \cite{LSPV}. The importance of
this object was already known in the context of the black hole analogy \cite{LSVP}, the
novelty here is the natural description in terms of three-qubit operators.

It is also interesting that keeping a certain set of $63$ lines of $\mathcal{G}_3$
one can obtain the spit Cayley hexagon of order two \cite{LSV}. Since keeping all points
and deleting lines weakens the condition of being a geometric hyperplane, all
hyperplanes of $\mathcal{G}_{3}$ can also be viewed as hyperplanes of the
hexagon, but the latter contains many more types of hyperplanes\cite{Frohardt}. From the
physical point of view, there are some hints that the Cayley hexagon might have
a role in understanding the connection between the three-qubit Pauli group and
the $E_{7(7)}$-symmetric entropy formula of black holes in $N=8$ $D=4$
supergravity.

We also have two special geometric hyperplanes in $\mathcal{G}_n$ for any $n$
if we fix the representation of the $n$-qubit Pauli group as tensor products of
the usual Pauli matrices. These are $H_{0}$ and $H_{YY\ldots Y}$. As it was
already mentioned, the first one consists of Pauli operators with symmetric
matrices as representatives. The latter one contains the operators built from
an even number of nontrivial (i.e. $X$, $Y$, or $Z$) matrices. These $n$-qubit
operators are exactly the self-dual ones with respect to the Wootters spin-flip \cite{Wootters}
transformation (apart from the identity matrix).

For $n=3$ these Wootters self-dual Pauli operators form an $H$-hyperplane giving rise to a $GQ(2,4)$ underlying the geometry of the $E_{6(6)}$ symmetric entropy formula for black holes and black strings.
Of course this hyperplane is just one from the aforementioned  $28$ 
possible ones with Arf invariant equals to 1. They also have the structure of a $GQ(2,4)$. Clearly, all of these hyperplanes can be used to describe the same $E_{6(6)}$ symmetric black hole and black string entropy formula, but with the points having different noncommutative labellings. 
This situation can be regarded as the finite geometric analogue 
of the standard usage of different local coordinates for the underlying manifold in (pseudo) Riemannian geometry.
As also emphasized in our recent paper \cite{LSVP}, the important novelty here is the intrinsically noncommutative nature of these coordinates. Using the unified framework as developed in this paper the mathematical and physical implications of these observations 
are certainly worth exploring further. 

\section{Conclusion}
We have associated a point-line incidence geometry to every $n$-qubit generalized
Pauli group, from which the group can fully be recovered. This contains points
and lines of a symplectic polar space of rank $n$ and order $2$ which describes
the commutation relations of the Pauli group \cite{Saniga1}.

For $n\ge 3$ this incidence structure has a Veldkamp space in the stronger sense
which enables us to identify distinguished subsets of the group indepentently
from its representation. These are the geometric hyperplanes (Veldkamp points)
and the intersections of pairs of hyperplanes (Veldkamp lines).

This formalism also creates a nice unifying picture of finite geometric results
in connection with the black hole analogy. Namely, the generalized quadrangle
with $(2,4)$ parameters, which is intimately connected to the $E_{6(6)}$-symmetric
black hole entropy formula \cite{LSVP}, and the split Cayley hexagon of order two, which
is related to the $E_{7(7)}$-symmetric black hole entropy formula in $N=8$ $D=4$
supergravity \cite{LSV} both can be found as a subgeometry in the incidence structure
describing the Pauli group of three qubits.

\end{document}